\documentclass[sn-basic]{sn-jnl}

\usepackage{graphicx}%
\usepackage{multirow}%
\usepackage{amsmath,amssymb,amsfonts}%
\usepackage{amsthm}%
\usepackage{mathrsfs}%
\usepackage[title]{appendix}%
\usepackage{xcolor}%
\usepackage{textcomp}%
\usepackage{manyfoot}%
\usepackage{booktabs}%
\usepackage{algorithm}%
\usepackage{algorithmicx}%
\usepackage{algpseudocode}%
\usepackage{listings}%

\usepackage{comment}
\newcommand{\heading}[1]{\noindent\textbf{#1}}

\newcommand{\colred}[1]{{\textcolor{red}{#1}}}
\newcommand{\colyellow}[1]{{\textcolor{orange}{#1}}}
\newcommand{\colgreen}[1]{{\textcolor{green}{#1}}}

\usepackage{hyperref}

\theoremstyle{thmstyleone}%
\theoremstyle{thmstyletwo}%

\theoremstyle{thmstylethree}%

\begin{document}

\title[Neoplasms diagnosis via ECG]{Explainable machine learning for neoplasms diagnosis via electrocardiograms: an externally validated study}


\author[1]{\fnm{Juan Miguel} \sur{Lopez Alcaraz}}\email{juan.lopez.alcaraz@uol.de}

\author[2]{\fnm{Wilhelm} \sur{Haverkamp}}\email{wilhelm.haverkamp@dhzc-charite.de}

\author*[1]{\fnm{Nils} \sur{Strodthoff}}\email{nils.strodthoff@uol.de}

\affil[1]{\orgdiv{AI4Health Division}, \orgname{Carl von Ossietzky Universität Oldenburg}, \orgaddress{\street{Ammerländer Heerstraße 114-118}, \city{Oldenburg}, \postcode{26129}, \state{Lower Saxony}, \country{Germany}}}

\affil[2]{\orgdiv{Department of Cardiology, Angiology and Intensive Care Medicine}, \orgname{Charité Campus Mitte, German Heart Center of the Charité-University Medicine}, \orgaddress{\street{Augustenburger Pl. 1}, \city{Berlin}, \postcode{13353}, \state{Berlin}, \country{Germany}}}


\abstract{

\textbf{Background:} Neoplasms are a major cause of mortality globally, where early diagnosis is essential for improving outcomes. Current diagnostic methods are often invasive, expensive, and inaccessible in resource-limited settings. This study explores the potential of electrocardiogram (ECG) data, a widely available and non-invasive tool for diagnosing neoplasms through cardiovascular changes linked to neoplastic presence.

\textbf{Methods:} A diagnostic pipeline combining tree-based machine learning models with Shapley value analysis for explainability was developed. The model was trained and internally validated on a large dataset and externally validated on an independent cohort to ensure robustness and generalizability. Key ECG features contributing to predictions were identified and analyzed.

\textbf{Results:} The model achieved high diagnostic accuracy in both internal testing and external validation cohorts. Shapley value analysis highlighted significant ECG features, including novel predictors. The approach is cost-effective, scalable, and suitable for resource-limited settings, offering insights into cardiovascular changes associated with neoplasms and their therapies.

\textbf{Conclusions:} This study demonstrates the feasibility of using ECG signals and machine learning for non-invasive neoplasm diagnosis. By providing interpretable insights into cardio-neoplasm interactions, this method addresses gaps in diagnostics and supports integration into broader diagnostic and therapeutic frameworks.

}


\maketitle

\section{Background}

\subsection{Research objective}

Neoplasms are among the leading causes of death globally with a 2024 projection of over 2 millon new neoplasms cases and more than 600.000 neoplasms deaths in the United States alone \cite{siegel2024cancer}. Despite progress in medical diagnostics and treatments, timely diagnosis continues to pose a significant challenge, as many neoplasms are identified only at advanced stages. Such delays adversely affect survival rates, highlighting the pressing need for accessible, non-invasive, and cost-effective diagnostic methods \citep{fitzgerald2022future}. Current diagnostic methods, including imaging, biopsies, and tumor biomarkers, are often invasive, resource-intensive, or inaccessible in low-resource settings \citep{crosby2022early}. These limitations highlight the necessity for innovative approaches to improve neoplasms detection and outcomes.

Electrocardiograms (ECGs), long regarded as a cornerstone for diagnosing cardiovascular conditions, have shown promise beyond their traditional applications. By capturing the heart's electrical activity, ECGs provide critical insights into cardiac rhythm and function. Recent advances have expanded their utility into non-cardiac domains, such as predicting laboratory value abnormalities \citep{alcaraz2024cardiolablaboratoryvaluesestimation}, patient deterioration in emergency settings \citep{alcaraz2024mds}, and other systemic health indicators, as reviewed in \citep{topol2021s}. These studies suggest that the ECG, in combination with machine learning methods, could play an important role in identifying broader physiological disruptions.

The relationship between neoplasms and the cardiovascular system is well-documented, particularly in the emerging field of cardio-oncology. Neoplasms can induce subtle cardiac alterations detectable through the ECG, whether by direct invasion, paraneoplastic syndromes, or systemic effects such as inflammation and hypercoagulability \citep{ogilvie2024cardiac}. Additionally, neoplasms therapies, including chemotherapy and targeted treatments, are associated with cardiotoxicity, which may lead to arrhythmias, ischemia, or myocardial dysfunction \citep{herrmann2020adverse}. Despite these known associations, the ECG remains underutilized as a diagnostic tool for neoplasms. Nevertheless, the ability to detect malignancy-related patterns in ECG signals offers a compelling opportunity to enhance neoplasms diagnosis.

This study investigates the integration of ECG features with demographic data to improve neoplasm diagnoses using tree-based machine learning models. The objective is to develop an accessible, non-invasive, and interpretable diagnostic tool to aid in detection and monitoring of neoplasms. By complementing existing diagnostic methods and addressing their shortcomings, this approach aims to enhance neoplasm outcomes and expand access to diagnostic solutions.

\subsection{Literature review}

\subsubsection{Overview of neoplasms diagnoses}
Traditional diagnostic approaches for neoplasms rely heavily on serum biomarkers, imaging techniques, and tissue biopsies. Serum biomarkers, while offering a less invasive alternative through blood sampling, often suffer from limited sensitivity and specificity, especially for initial stages of neoplasms or tumors located in hard-to-reach anatomical areas \citep{srinivas2001trends}. Imaging modalities such as CT scans, MRIs, and PET scans are essential for detecting and staging neoplasms but apart of being unaccesible for large population groups \citep{dosanjh2024access}, are resource-intensive and may not always distinguish between benign and malignant lesions with high accuracy. Tissue biopsies, considered as the gold standard for diagnosing neoplasms, are invasive procedures that carry risks such as bleeding, infection, and sampling errors, which can lead to misdiagnoses or delays in treatment. These challenges highlight the need for advanced diagnostic tools that are truly non-invasive, improving timely detection, reducing procedural risks, and supporting personalized treatment strategies.

\subsubsection{ECG as a diagnostic tool}
Electrocardiograms (ECG) play an important role in diagnosing and monitoring cardiovascular diseases, providing a non-invasive means to evaluate the heart's electrical activity. Traditionally, ECG analysis has focused on detecting arrhythmias, myocardial infarctions, and other cardiac disorders through electrical signal patterns. However, recent advances have broadened its applications beyond cardiology, as highlighted in reviews such as \citep{topol2021s,siontis2021artificial}, with studies demonstrating its potential for systemic health monitoring. For example, \citet{strodthoff2024prospects} recently showcased the ability to predict a wide range of cardiac and non-cardiac neoplasms from a single ECG from a unified model. Given its non-invasive nature, affordability, and accessibility, ECG emerges as a promising tool for developing novel diagnostic models, including those targeting neoplasm-related conditions.

\subsubsection{Cardiovascular-neoplasms interactions}
The interactions between the cardiovascular system and neoplasms are complex and multifaceted, with neoplasms influencing cardiovascular health and vice versa. Certain neoplasms, such as those of the lung and breast, are associated with increased risks of cardiovascular complications due to tumor-induced hypercoagulability, leading to thromboembolic events like deep vein thrombosis and pulmonary embolism \citep{ogilvie2024cardiac}. Neoplasms survivors, including those treated for childhood neoplasms, also face an increased risk of cardiovascular issues later in life \citep{hammoud2024burden}. Additionally, neoplasms treatments, including chemotherapy, radiotherapy, and targeted therapies, frequently induce cardiotoxicity, manifesting as neoplasms like heart failure, arrhythmias, and myocardial ischemia \citep{herrmann2020adverse,altena2009cardiovascular}. Conversely, cardiovascular conditions can affect neoplasms progression and outcomes. Chronic heart diseases, through mechanisms like reduced systemic perfusion and hypoxia, may create a microenvironment conducive to tumor growth and metastasis. Furthermore, shared risk factors, including obesity, smoking, and systemic inflammation, exacerbate both cardiovascular and oncological neoplasms, underscoring their interconnected pathophysiology \citep{herrmann2014evaluation}. These bidirectional relationships highlight the importance of integrated multidimensional approaches for diagnosing, managing, and preventing cardiovascular complications in oncology and vice versa.

\subsection{ECG in oncology}

Cardio-oncology is an emerging multidisciplinary field that addresses the cardiovascular health of patients with cancer \citep{10.1093/ehjci/jeac106}. 
With improved cancer survival rates and the increasing use of cardiotoxic therapies, there is a growing need to understand, detect, and manage cardiac complications in this population.
Electrocardiography (ECG) plays a critical role in this setting, offering a readily accessible tool for early detection of arrhythmias, myocardial injury, and conduction disturbances.
Cancer patients are at increased risk for arrhythmias and other ECG abnormalities due to a range of factors, including direct tumor effects (e.g., cardiac infiltration or compression), therapy-induced cardiotoxicity (e.g., chemotherapy, radiotherapy, immunotherapy), paraneoplastic syndromes and immune-mediated inflammation, and a possible inherent pro-arrhythmic state, even before treatment initiation \citep{khera2025artificial}.

Recent literature has expanded our understanding of ECG changes in cancer patients: Case reports illustrate how cardiac metastases can imitate acute coronary syndromes (ACS). For instance, ST-segment elevation in a lung cancer patient was due to right ventricular metastasis, despite normal cardiac biomarkers \citep{samaras2007infarction}. A systematic review of 36 reports found that cardiac metastases often produce convex ST elevations in specific coronary territories without typical ischemic progression \citep{akgun2020electrocardiographic}. Mechanical effects also contribute. In a cohort of 264 lung cancer patients, the presence of J waves correlated strongly with direct tumor-heart contact \citep{hayashi2017relationship}. Paraneoplastic and immunerelated mechanisms can also alter ECGs. One patient on immune checkpoint inhibitors developed ECG findings suggestive of myocarditis alongside immune-mediated myositis \citep{xu2022ominous}. Even before treatment, cancer patients may show abnormal ECGs. A propensity-matched study comparing newly diagnosed cancer patients with surgical controls found significantly more baseline conduction delays and repolarization abnormalities in the cancer group \citep{golemi2023baseline}, suggesting a possible inherent pro-arrhythmic state.

Overall, the ECG remains a frontline diagnostic tool in cardio-oncology. Understanding its nuances in cancer patients—across various stages of disease and treatment is essential for risk stratification, monitoring, and timely intervention \citep{pohl2021ecg,flore2023mechanisms,wright2025arrhythmia}. As the field evolves, integrating ECG findings with imaging, biomarkers, and genetic data will further enhance cardiovascular care in oncology.

\section{Methods}

\subsection{Dataset}

\begin{figure*}[ht!]
    \centering
    \includegraphics[width=\textwidth]{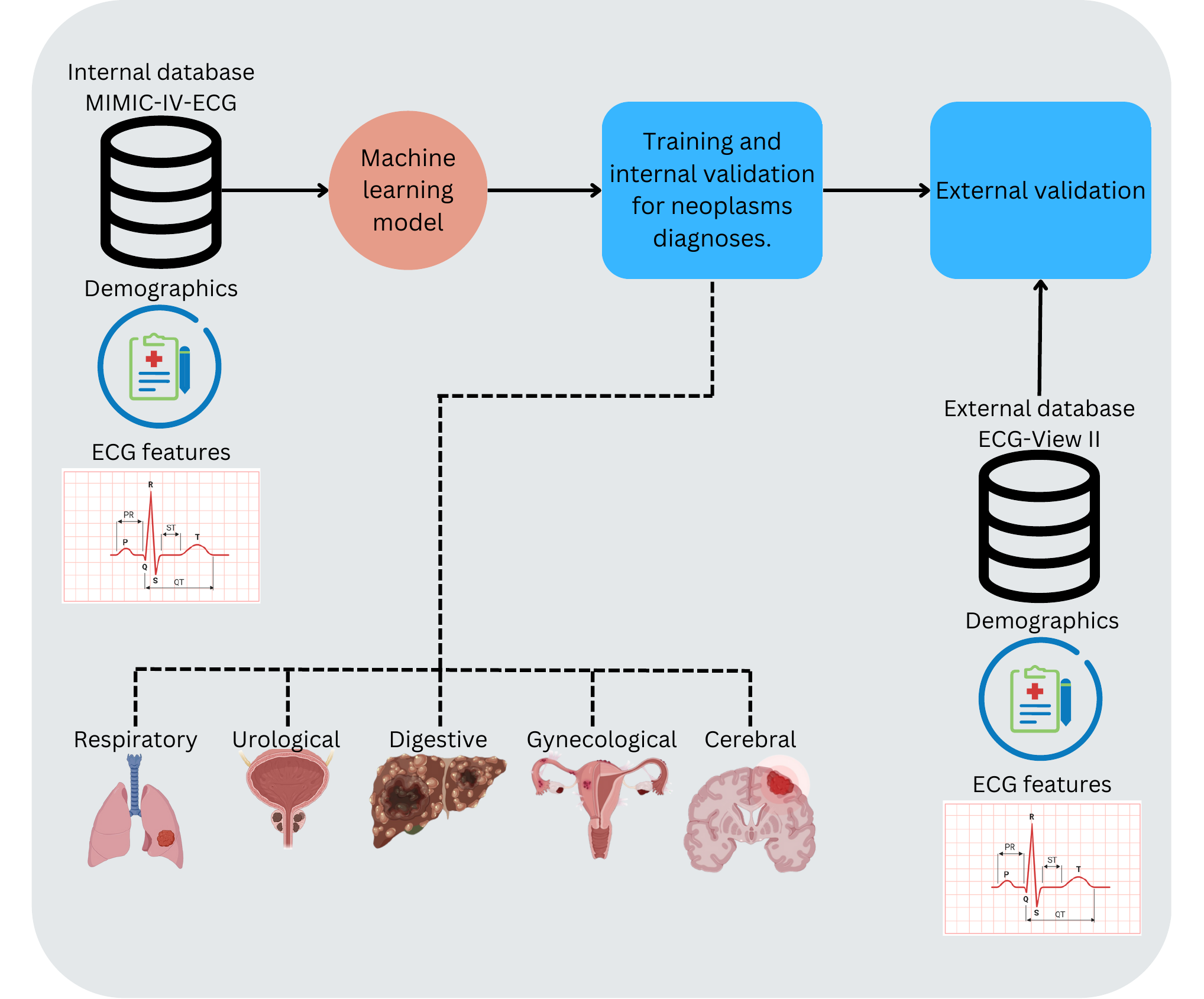}
    \caption{Schematic representation of our proposed approach. We use as internal dataset the MIMIC-IV-ECG dataset from which we use as input features demographics and ECG features to train a tree-based model and diagnose diverse neoplasms. For external validation we take a second cohort of patients from the ECG-View II dataet from which we collect the same set of features and neoplasms targets. The definition of neoplasms are represented by ICD10-CM codes.}
    \label{fig:abstract}
\end{figure*}

\begin{table}[ht!]
    \centering
    \caption{A summary of variable characteristics across samples, including demographic details such as gender counts (with ratios) and the median age in years (with interquartile range, IQR), along with age distribution represented by quantile ratios. Similarly, it covers the median (IQR) values for ECG features like the RR-interval, PR-interval, QRS-duration, QT-interval, and QTc-interval in milliseconds, as well as the P-wave axis, QRS axis, and T-wave axis in degrees.}
    \begin{tabular}{lll}
    \hline
    \textbf{Variable} & \textbf{MIMIC-IV-ECG} & \textbf{ECG-View II} \\ \hline\hline
    \textbf{Gender (\%)} &  &  \\ \hline
    Female     & 226,892 (48.50)  & 375,733 (48.44) \\
    Male   & 240,837 (51.49)  & 399,802 (51.55) \\ \hline
    \textbf{Age (\%)} &  &  \\ \hline
    Median years (IQR) & 66 (25) & 52 (25) \\ 
    Quantile 1 & 18-53 (23.83) & 18-40 (24.03) \\
    Quantile 2 & 53-66 (25.16) & 40-52 (25.75) \\
    Quantile 3 & 66-78 (25.60) & 52-65 (24.94) \\
    Quantile 4 & 78-101 (25.40) & 65-109 (25.28) \\ \hline
    \textbf{ECG features (IQR)} &  &  \\ \hline
    RR-interval & 769 (264)  & 857 (227) \\
    PR-interval & 158 (38)  & 158 (28) \\
    QRS-duration & 94 (23)  & 90 (14) \\
    QT-interval & 394 (68)  & 392 (48) \\
    QTc-interval & 447 (47)  & 421 (37) \\
    P-wave-axis & 51 (32) & 53 (28) \\
    QRS-axis & 13 (61) & 48 (49) \\
    T-wave-axis & 42 (58)  & 44 (33) \\ \hline
    \end{tabular}%
    \label{tab:descriptive}
\end{table}

Our primary dataset for training and internal evaluation was derived from the MIMIC-IV-ECG database \citep{johnson2023mimic,MIMICIVECG2023}, a subset of a large-scale critical care dataset collected at the Beth Israel Deaconess Medical Center in Boston, Massachusetts. This dataset encompasses patients admitted to the emergency department (ED) and intensive care unit (ICU). Target variables are based on discharge diagnoses encoded using the International Classification of Diseases Clinical Modification (ICD-10-CM). While a wide range of neoplasm-related codes is explored, this study focuses on those achieving internal and external validation AUROC scores above 0.7, covering neoplasms across diverse physiological systems.

To construct a comprehensive and harmonized feature set, ECG features from MIMIC-IV were aligned with those from the ECG-VIEW-II database \citep{kim2017ecg}, which serves as our secondary dataset for external validation. ECG-VIEW-II includes data collected from patients at a South Korean tertiary teaching hospital. The standardized feature set comprises ECG-derived measurements (RR-interval, PR-interval, QRS-duration, QT-interval, QTc-interval in milliseconds; P-wave-axis, QRS-axis, and T-wave-axis in degrees) alongside demographic attributes (binary sex and age as a continuous variable).

For the internal dataset, stratified folds are created based on diagnoses, age, and gender distributions, utilizing an 18:1:1 split as described in prior work \citep{strodthoff2024prospects}. A comparable stratification procedure is applied to the external dataset to maintain consistency. The training process prioritizes MIMIC-IV-ECG due to its broader ethnic diversity compared to ECG-VIEW-II, thereby enhancing the model's generalization across diverse populations,as demonstrated in previous research \citep{alcaraz2024estimationcardiacnoncardiacdiagnosis}, which employs a similar approach mostly for cardiac conditions and \citep{alcaraz2024electrocardiogrambaseddiagnosisliverdiseases} for diverse liver disease conditions. This approach ensures robust internal training and reliable external validation across ethnically and geographically distinct cohorts.

\subsection{Prediction models}
In this study, we develop individual tree-based models using Extreme Gradient Boosting (XGBoost) to address binary classification tasks, with a separate model for each selected ICD-10-CM code. To prevent overfitting, we implement early stopping with a patience of 10 iterations on the validation fold during training. To this end, model performance is evaluated using the area under the receiver operating characteristic curve (AUROC) on the test fold internally, and the complete external dataset as external evaluation. In addition to XGBoost, we include logistic regression (LR), and a multi-layer perceptron (MLP) as baseline models to contextualize performance as well as their computational complexity. Based on the results of this model benchmark, XGBoost was selected as the primary model throughout the manuscript. Detailed benchmarking results and hyperparameter settings for all models are provided in the appendix. To improve calibration, we apply model-agnostic calibration and fit isotonic regression models on the validation set and report calibrated test set results.

\subsection{Evaluation procedure}
A recent review on evaluation criteria for prediction algorithms \citep{vancalster2024performanceevaluationpredictiveai} identified three evaluation categories for predictive medical AI models: discrimination, calibration, and clinical utility. We address discriminative performance in terms of AUROC scores evaluated both on the internal test set and on an external dataset along with 95\% confidence intervals derived through empirical bootstrapping with 1000 iterations. To address calibration, we show calibration curves for the internal test set. Finally, we  demonstrate clinical utility through a net benefit analysis in comparison to common baselines ("refer all" and "refer none") via decision curve analysis \citep{vickers2006decision}.

\subsection{Explainability}
Our goal extends beyond simply evaluating model performance. In order to gain deeper insights into the trained models, we incorporate Shapley values into our workflow \citep{lundberg2020local}. These values offer a way to assess feature importance by quantifying the individual contribution of each feature to the model's predictions. The computational complexity and hyperparemeters setting for this approach are given in the appendix.

\section{Results}

\subsection{Predictive performance}

\begin{table*}[ht!]
    \centering
    \caption{Predictive performance results for the investigated neoplasms of diverse physiological systems. We list internal (MIMIC-IV) and external (ECG-View) AUROC performances with 95\% confidence intervals as well as the class prevalance of the conditon in each dataset.}
    \resizebox{0.98\textwidth}{!}{%
    \begin{tabular}{lll}
    \hline
    \textbf{Code: Description} & \textbf{Internal AUROC (95\% CI) [Prev.]} & \textbf{External AUROC (95\% CI) [Prev.]} \\ \hline\hline
    \textbf{Respiratory neoplasms} &  &  \\ \hline
    C34: Lung cancer  & 0.800 (0.784, 0.815) [1.61\%] & 0.767 (0.765, 0.771) [1.83\%] \\
    C341: Upper lung cancer &  0.723 (0.706, 0.750) [0.48\%]  &  0.738 (0.723, 0.751) [0.04\%] \\
    C343: Lower lung cancer & 0.855 (0.788, 0.887) [0.27\%]  & 0.752 (0.747, 0.770) [0.04\%] \\
    C349: Unspecified lung cancer &   0.792 (0.755, 0.822) [0.57\%] &  0.759 (0.757, 0.760) [1.72\%] \\ \hline

    \textbf{Urological neoplasms} &  &  \\ \hline
    C61: Prostate cancer & 0.756 (0.746, 0.781) [1.27\%] &  0.795 (0.792, 0.797) [0.8\%] \\
    N40: Benign prostatic hyperplasia (BPH) &  0.749 (0.739, 0.760) [12.38\%] &  0.820 (0.817, 0.821) [1.0\%] \\
    N400: BPH without symptoms &  0.739 (0.727, 0.751) [9.55\%] &  0.828 (0.823, 0.831) [0.87\%] \\
    C679: Bladder cancer, unspecified &  0.833 (0.803, 0.893) [0.18\%] & 0.757 (0.748, 0.762) [0.33\%] \\ \hline
         
    \textbf{Digestive neoplasms} &  &  \\ \hline
    C15: Esophageal cancer &  0.818 (0.780, 0.876) [0.18\%] & 0.810 (0.807, 0.815) [0.26\%] \\
    C22: Liver cancer & 0.808 (0.782, 0.825) [0.59\%] &  0.719 (0.715, 0.722) [1.43\%] \\
    C24: Biliary tract cancer  & 0.837 (0.756, 0.904) [0.11\%] &  0.706 (0.702, 0.712) [0.26\%] \\ \hline

    \textbf{Gynecological neoplasms} &  &  \\ \hline
    D25: Leiomyoma of uterus & 0.808 (0.735, 0.854) [0.52\%] & 0.730 (0.727, 0.732) [3.26\%] \\
    N80: Endometriosis  & 0.879 (0.845, 0.907) [0.16\%] &  0.753 (0.750, 0.757) [1.47\%] \\ \hline

    \textbf{Cerebral neoplasms} &  &  \\ \hline
    C793: Brain metastases & 0.738 (0.712, 0.762) [0.71\%] & 0.699 (0.693, 0.707) [0.13\%] \\ \hline
    \end{tabular}
    }
    \label{tab:aurocs}
\end{table*}

Table \ref{tab:aurocs} shows the predictive performance of our model across multiple neoplasms, assessed through AUROC scores on the internal and external test sets. The 95\% prediction intervals offer an understanding of the reliability of these metrics. Similarly, within each figure we report the class prevalance of each neoplasm within their respective datasets, which provides context in regards the representative distribution of the populations. The MIMIC cohort shows prevalence between 0.11\% to 12.38\%, whereas the Korean cohort shows significantly lower prevalences between 0.04\% to 3.26\%. 

Notably, from the respiratory system, the most accurately predictable neoplasm is ``C343: Lower lung cancer'' with 0.855 AUROC, from the urological system ``C679: Bladder cancer, unspecified'' with 0.833 AUROC, from the digestive system ``C24: Biliary tract cancer'' with 0.837 AUROC, from the gynecological system ``N80: Endometriosis'' with 0.879, and from the cerebral system ``C793: Brain metastases'' with 0.738. For simplicity, we restrict ourselves to results achieved by the XGBoost model. In the appendix, we present additional results for the LR and MLP baselines. All three models often perform comparably, which underlines the robustness of our findings. Across all tasks, the XGBoost model shows the overall best performance and was therefore selected for all further investigations.

\begin{figure*}[ht!]
    \centering
    \includegraphics[width=\textwidth]{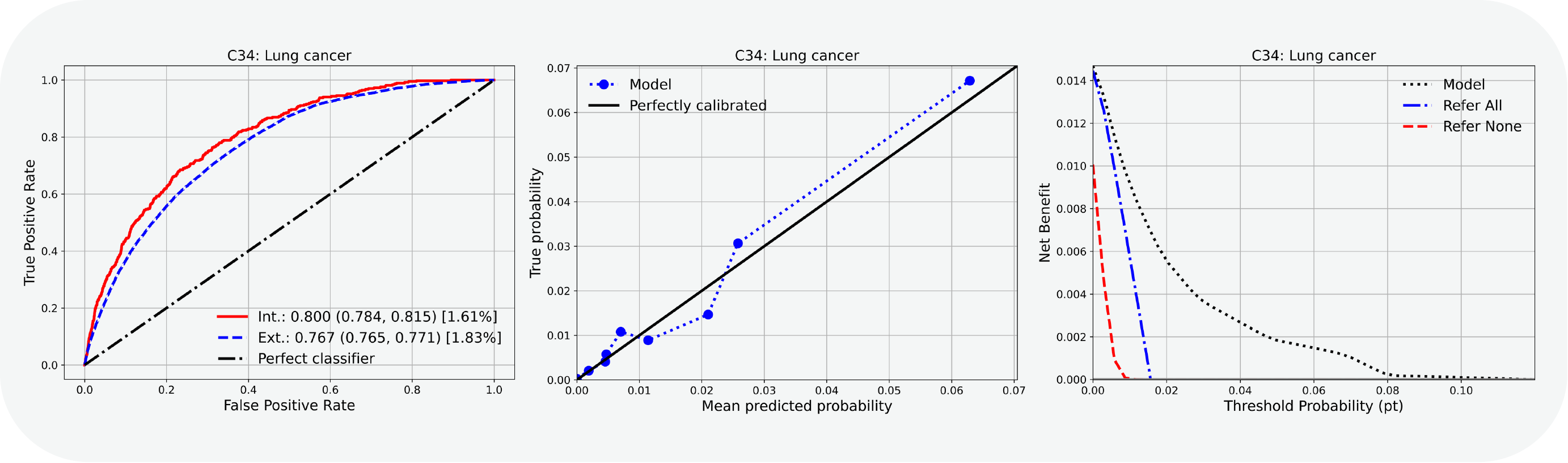}
    \caption{Exemplary performance analysis for the condition ``C34: Lung cancer'' condition, showing the model's performance across three key evaluation metrics: AUROC curves (discrimination), calibration curves (agreement between predicted and observed risks), decision curve analysis (net benefit compared to "refer all" and "refer none" strategies). Corresponding plots for all other considered conditions can be found in the appendix.}

    \label{fig:main_grid}
\end{figure*}

Extending beyond discriminative performance in terms of AUROC scores, we demonstrate three facets of model performance in Figure~\ref{fig:main_grid} at the example of condition ``C38: Lung cancer''. The ROC curves (left panel) align with the high predictive performance in both the internal and external test set. The calibration curve (middle panel) demonstrates good calibration, underlining the reliability of predicted probabilities. The decision curve (right panel) demonstrates clinical utility in comparison to the two baseline strategies considered. Given the low prevalences of all conditions in the dataset, both the calibration curves and the relevant part of the decision curves concentrate in the low probability threshold range.

\subsection{Explainability}

\begin{figure*}[ht!]
    \centering
    \includegraphics[width=\textwidth]{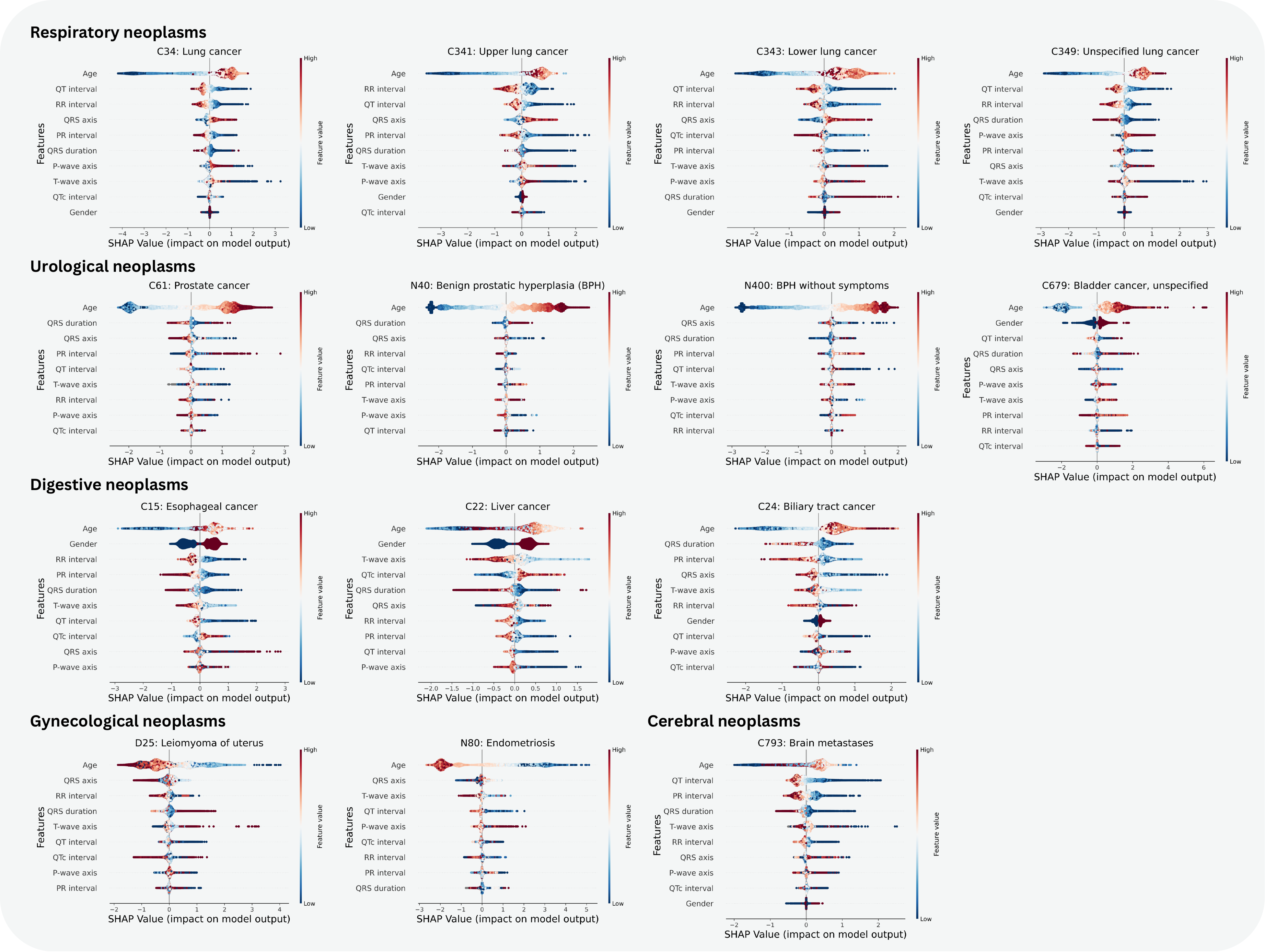}
    \caption{Explainability results for the investigated neoplasms. The beeswarm plot visualizes through a single dot per feature and sample if the feature contributes positively (right hand side) or negatively (left hand side) to the model prediction. In addition, the color-coding allows to infer if a point is associated with high (red) or low (blue) feature values.}
    \label{fig:shapley}
\end{figure*}

Figure \ref{fig:shapley} presents the explainability results using Shapley values. Across all investigated neoplasms, age is the most important feature. Higher age values (older patients) contribute positively to the respiratory, urological, and digestive systems. In contrast, lower age values (younger patients) contribute positively to the gynecological system. The cerebral system exhibits a mix of contributions from both younger and older patients. Similarly, low QT-interval values (faster ventricular repolarization) contribute positively across all investigated neoplasms, except in cases of malignant neoplasm of the bladder, which show only a few cases with high values. Apart from gender-specific neoplasms, male sex contribute more positively overall than female sex.

\heading{Respiratory}
For the investigated respiratory neoplasms, QT-interval and RR-interval represent the two most important ECG features. In terms of feature values, low values of the RR-interval (faster heart rates), PR-interval (shorter atrial conduction time), and QRS-duration (more efficient ventricular depolarization) generally contribute positively, with the exception of lower lung cancer, where high QRS-duration values (delayed ventricular conduction) are more significant. High values of the QRS axis (altered electrical orientation of the ventricles) also contribute positively across these neoplasms.

\heading{Urological}
For the investigated urological neoplasms, QRS-duration and QRS axis are the two most important ECG features. In terms of feature importance, low values of the QRS axis (altered electrical orientation of the ventricles) and P wave axis (abnormal atrial electrical orientation) contribute positively.

\heading{Digestive}
For the investigated digestive neoplasms, male sex is the most important feature for esophageal and liver cancer. Male sex also contribute more than female sex for the biliary tract, albeit in a less pronounced fashion. There is no consistent ECG feature of high importance across all the investigated neoplasms of the system. In terms of feature value, low values of the PR-interval (indicating faster atrial conduction), QRS-duration (shorter ventricular depolarization time), T-wave-axis (altered repolarization pattern), and QT-interval (faster ventricular repolarization) contribute positively.

\heading{Gynecological}
For the investigated gynecological neoplasms, the QRS axis is the most important ECG feature. In terms of feature value importance, low values of the QT-interval (faster ventricular repolarization) contribute positively, suggesting a quicker recovery of the ventricles after each heartbeat.

\heading{Cerebral}
For the only investigated cerebral neoplasm, the most important ECG features are the QT-interval, PR-interval, QRS-duration, T-wave-axis, and RR-interval, in that order. In terms of feature value importance, low values of all of these contribute positively. Low QT-interval values (faster ventricular repolarization), low PR-interval values (shorter atrioventricular conduction), low QRS-duration (faster ventricular depolarization), low T-wave-axis values (potentially indicating quicker repolarization of the ventricles), and low RR-interval values (indicating faster heart rate), which suggest stress response and systematic inflammation associated with neoplasms.

Finally, we include in the appendix a comparison of ECG features summarized using the median and interquartile range across binary outcomes (diagnosed vs. not diagnosed). This analysis supports the validity of our approach and highlights clinically meaningful distinctions in ECG characteristics between the two groups.

\section{Discussion}

\subsection*{ECG biomarkers for non-cardiovascular conditions}
Detecting neoplasms through ECG features may initially seem unconventional, as the ECG is traditionally associated with diagnosing cardiovascular conditions. However, the physiological interplay between the cardiovascular system and neoplastic processes offers a unique perspective for diagnostic innovation. Although the mechanisms linking neoplasms to ECG abnormalities are not yet fully understood, they present an intriguing avenue for further investigation. Our findings uncover specific ECG patterns that serve as distinctive markers for neoplastic conditions, suggesting underlying physiological connections that are detectable through machine learning methods. This interdisciplinary approach underscores the potential of bridging oncology and cardiology to uncover novel diagnostic pathways and improve non-invasive neoplasms diagnosis strategies.

\subsection*{Predictive performance}

The remarkable predictive strength of a select group of ECG features emphasizes their capacity to accurately identify neoplasms from a single ECG. Consistently high AUROC values in both internal and external validations confirm the robustness of these features, even across varied cohorts. The unique patterns identified across different physiological systems highlight the interconnectedness between cardiac and oncological health. Remarkably, our approach is able to distinguish between benign and malignant neoplasms or diverse neoplasms with alike symptoms such as malignant neoplasm of prostate against benign prostatic hyperplasia, as well as leiomyoma of the uterus and endometriosis.

The variation in predictive performance observed across different neoplasm types likely reflects underlying physiological and pathophysiological heterogeneity in how various cancers influence cardiac electrophysiology, as captured by the ECG. For example, neoplasms such as lower lung cancer due to their anatomical proximity to the heart or their potential to trigger paraneoplastic syndromes, systemic inflammation, or changes in autonomic regulation, may induce more pronounced alterations in ECG signals. These changes make such neoplasms more readily detectable by ECG-based models. Conversely, cancers that are located further from the thoracic cavity or that exert limited systemic effects may not manifest discernible ECG signatures, resulting in reduced model performance for those categories.

\subsection*{Feature importance}
In this study, age was identified as a key factor, with older patients contributing more to the most of the neoplasms except patients associated with gynecological neoplasms. This aligns with previous findings that report an increased incidence of ventricular arrhythmias linked with a worse prognosis in older neoplasms patients \citep{anker2021ventricular,albrecht2021spontaneous}. Additionally, our findings show that males contribute more than females across many neoplasms, which is consistent with studies showing a higher occurrence of premature ventricular contractions in male neoplasms patients \citep{anker2021ventricular}. Lastly, the association of lower QT-interval values across several neoplasms types mirrors findings that higher heart rates, as seen in tachycardia, are independent predictors of poor survival in neoplasms patients \citep{anker2016resting}.

\subsection*{Potential innovations and applications}
ECG is a valuable tool for detecting electrical abnormalities; however, it cannot directly diagnose or localize neoplasms. Accurate detection and localization require additional imaging modalities, such as echocardiography or MRI. Therefore, at this stage, we consider ECG a preliminary screening tool that can help identify abnormalities but must be complemented by imaging techniques for definitive neoplasm assessment.

Changes in the ECG may serve as indicators for the presence of heart damage or abnormal heart activity by the prescence of diverse neoplasms in patient's body, thus supporting neoplasms diagnostic and risk stratification once counfounding addressed. Nevertheless, for cardiac monitoring in oncology patients ECGs can be integrated into comprehensive cardio-oncology management strategies, where they are used for monitoring the cardiotoxicity of neoplasms treatments. This includes regular ECG checks alongside imaging modalities and cardiac biomarkers such as troponins and NT-proBNP, which help assess treatment-related cardiovascular risks. By monitoring ECG patterns during therapy, especially for high-risk drugs, clinicians can early detect signs of cardiotoxicity. These findings are invaluable in guiding clinical decisions, such as adjusting drug dosages, initiating cardioprotective strategies, or providing early interventions to mitigate further heart damage \citep{10.1093/eurheartj/ehw211,10.1093/ehjci/jeac106}. Ultimately, this integrated approach helps balance the efficacy of neoplasms therapies with the safety of the heart, improving the overall quality of life for patients while maintaining treatment effectiveness.

\subsection*{Limitations and future work}

First, regarding patient stratification, we acknowledge that external variables may introduce confounding effects, such as newly identified diagnoses and preexiting conditions. Since the ICD-10 codes in the dataset reflex a mix of these, the model predictions may partially capture therapy-induced cardiac changes, such as cardiotoxic effects of treatment, rather than signals solely related to the neoplasm itself. Resolving this ambiguity is an important next step for follow-up studies. Second, it is worth noting that prior work \citep{strodthoff2024prospects} has investigated label correlations for the MIMIC-IV dataset and found no significant label correlations. This defutes the potential claim that models detect other conditions commonly co-occurring with neoplasms. This aligns with very well with \citep{golemi2023baseline}, which clearly demonstrates the feasibility of finding cardiac abnormalities in newly diagnosed cancer patients.

Many ECG changes are non-specific and may arise from non-neoplasmsous conditions, such as electrolyte imbalances or ischemic heart disease, making it difficult to attribute abnormal ECG patterns to neoplasms alone. Future research should investigate how ECG abnormalities vary across age groups and distinguish these from typical age-related ECG changes \citep{ott2024using}. Moreover, exploring the causal relationships between ECG patterns and neoplasms will be crucial \citep{alcaraz2024causalconceptts}. Studies focusing on raw ECG waveforms, including external validation, could further enhance diagnostic accuracy \citep{strodthoff2024prospects, alcaraz2024mds}. The potential of raw ECG waveforms to outperform traditional ECG features in diagnostic tasks underscores the importance of continuing to refine this method for better diagnostic precision.

\section{Conclusion}
This study demonstrates the potential of using ECG biomarkers for the early detection of neoplasms, offering a non-invasive, cost-effective diagnostic tool. By identifying specific ECG patterns linked to neoplastic conditions, we show how the integration of machine learning methods can bridge the gap between cardiology and oncology, uncovering novel pathways for diagnosis. The strong predictive performance and feature importance findings highlight the robustness of ECG features in distinguishing between various neoplasm types, even differentiating benign from malignant conditions. This research underscores the value of ECG in cardio-oncology, with the potential to aid in both neoplasm diagnosis and monitoring treatment-related cardiotoxicity. Despite the limitations, including potential confounding by therapy-related factors, this study paves the way for further investigation into ECG's diagnostic capacity. Future studies can refine these findings, enhancing the accuracy and application of ECG-based neoplasm detection, ultimately improving patient care by integrating ECG monitoring into broader clinical management strategies.

\section*{Declarations}

\begin{itemize}
\item Ethics approval and consent to participate: Not applicable.

\item Consent for publication: Not applicable.

\item Availability of data and materials: Code for dataset preprocessing and experimental replications can be found in a dedicated \href{https://github.com/AI4HealthUOL/CardioDiag}{code repository}.

\item Competing interests: The authors declare no competing interests.

\item Funding: No specific funding was received for this research.

\item Author contribution: JMLA, and NS conceived and designed the project. JMLA conducted the full experimental analyses, with NS supervising them, and WH providing critical revision of clinical intellectual content. JMLA produced the first draft, NS and WH revised it. All authors critically revised the content and approved the final version for publication.

\end{itemize}

\bibliography{sn-bibliography}

\clearpage

\appendix

\section{Predictive performance}

\subsection{AUROC}

\begin{figure}[ht]
    \centering
    \includegraphics[width=\textwidth]{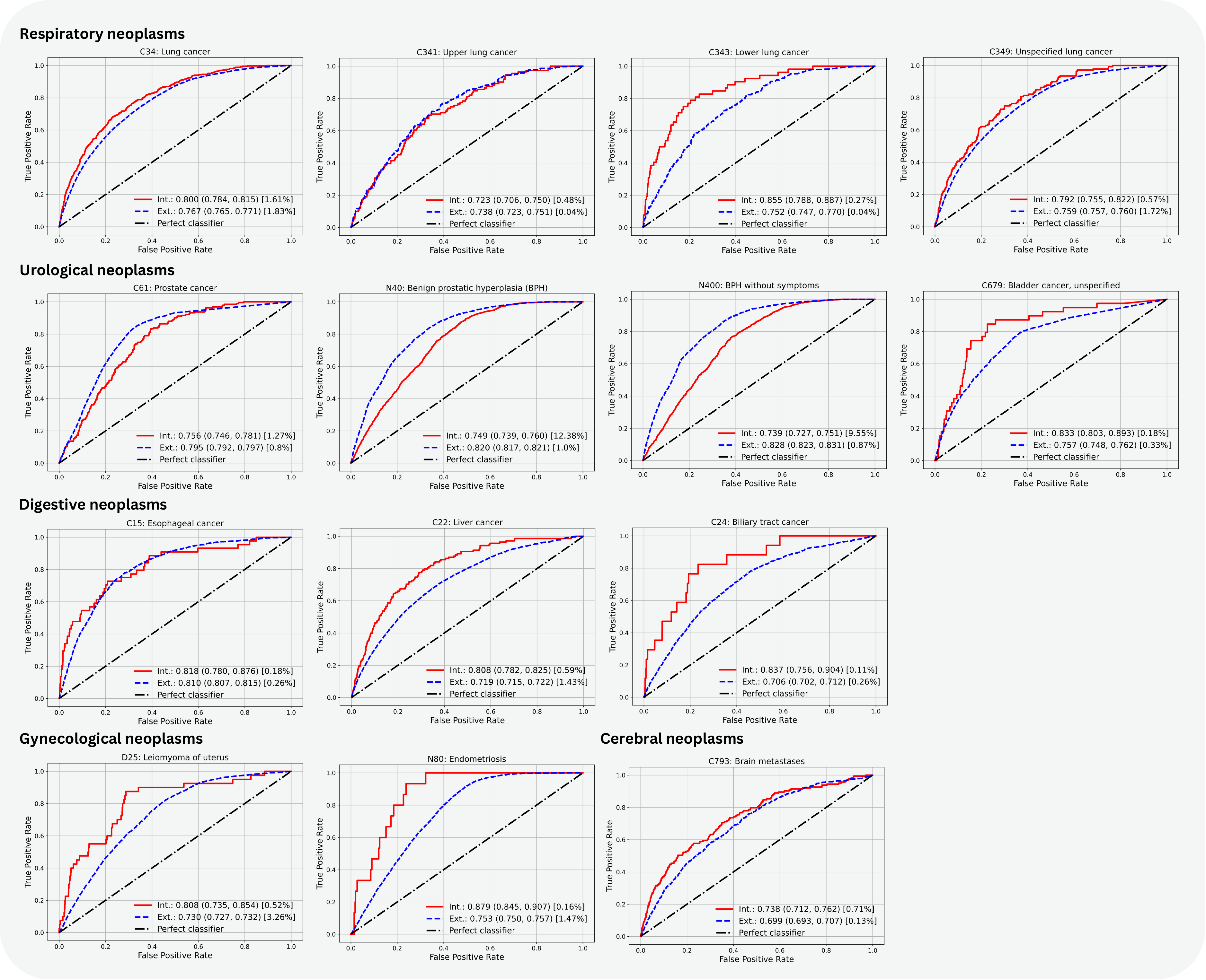}
    \caption{AUROC curves for all investigated labels, evaluating the model's ability to discriminate between positive and negative cases.}
    \label{fig:aurocs}
\end{figure}

Figure \ref{fig:aurocs} presents AUROC curves for all investigated diagnostic labels, including 95\% confidence intervals for both internal and external evaluations. The model demonstrates strong discriminative performance across all conditions. Importantly, AUROC values remain consistent between the internal (MIMIC-IV) and external (ECG-VIEW II) cohorts, with no substantial drop in performance. This indicates good generalizability and suggests the model maintains reliability when applied to independent populations.

\clearpage

\subsection{Calibration}

\begin{figure}[ht]
    \centering
    \includegraphics[width=\textwidth]{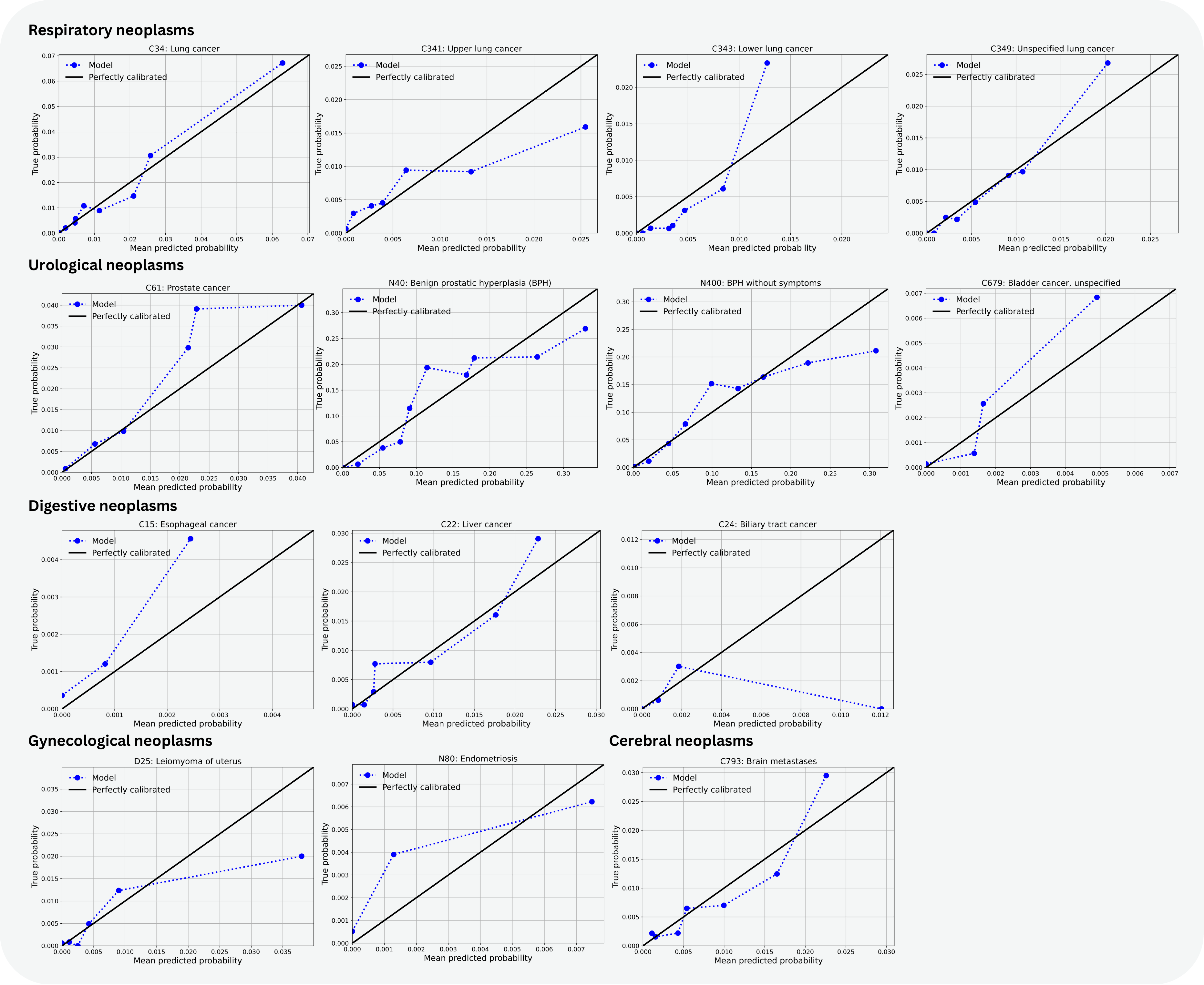}
    \caption{Calibration curves for each label, assessing the agreement between predicted probabilities and observed outcomes.}
    \label{fig:calibration}
\end{figure}

Figure \ref{fig:calibration} shows the calibration curves for each diagnostic label, evaluating the agreement between predicted probabilities and observed event rates. Overall, the models appear well calibrated, with predicted risks closely aligning with actual outcomes. We present the zoomed-in part of relevant probabilities based on low class prevalence, from where the lower half of the probability demonstrates particularly strong calibration. This indicates that within the actionable range of probabilities, the model provides reliable risk estimates that can support informed clinical decision-making.

\clearpage

\subsection{Net benefit}

\begin{figure}[ht]
    \centering
    \includegraphics[width=\textwidth]{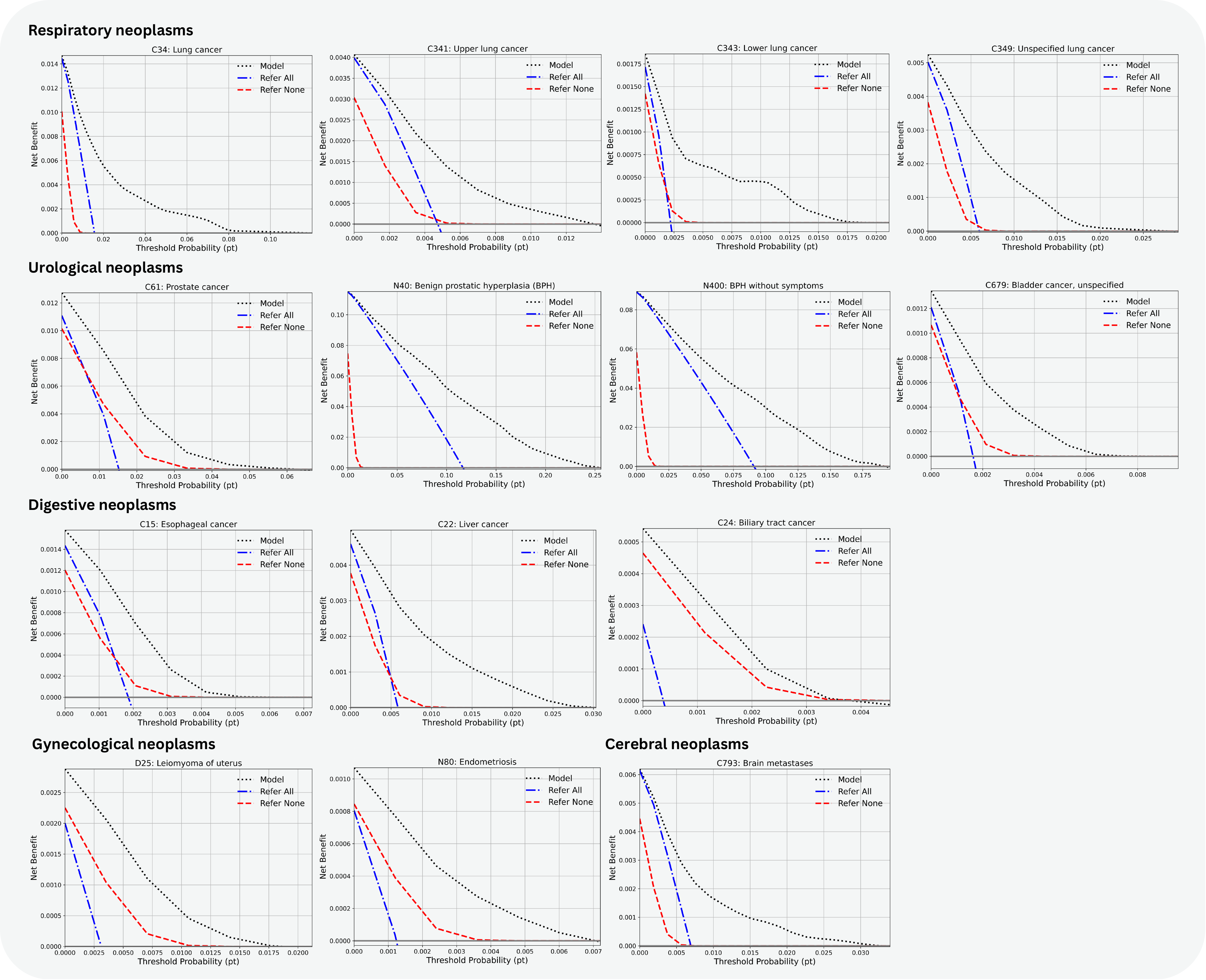}
    \caption{Decision curve analysis showing the net benefit of the model across thresholds, compared to "refer all" and "refer none" strategies for each label.}
    \label{fig:net_benefit}
\end{figure}

Figure \ref{fig:net_benefit} presents decision curve analyses for all investigated diagnoses, comparing the net benefit of our prediction model against two extreme strategies: referring all patients and referring none. Across all conditions, the model consistently demonstrates higher net benefit than both alternatives within clinically relevant threshold ranges. These thresholds lie predominantly in the lower probability range due to the low prevalence of positive cases, which is typical in population-wide screening or early detection settings.

\clearpage

\section{Computational complexity and hyperparameter settings}

All experiments were conducted on the high-performance computing (HPC) infrastructure of Carl von Ossietzky Universität Oldenburg. Each job was allocated 100 CPU cores and 100 GB of RAM. No GPUs were used; all models were trained exclusively on CPU-based resources. To ensure consistent data preprocessing across all models, we imputed missing values using the median computed from the training set within each fold. Additionally, continuous features (all except gender) were standardized, but only for models sensitive to feature scale and outliers, such as logistic regression and the multi-layer perceptron.

\subsection{Main classifier: XGBoost}

XGBoost was chosen as the primary model due to its robustness and performance on structured data. The model was configured for binary classification. Only a few hyperparameters were explicitly set; all others were left at their default values:

\begin{itemize}
    \item \texttt{objective}: \texttt{binary:logistic}
    \item \texttt{eval\_metric}: \texttt{logloss}
    \item \texttt{enable\_categorical}: \texttt{False}
    \item \texttt{use\_label\_encoder}: \texttt{False}
\end{itemize}

The remaining parameters (e.g., \texttt{max\_depth}, \texttt{learning\_rate}, \texttt{n\_estimators}, etc.) were kept at default settings provided by the XGBoost library (version 3.0.2).

\subsection{Baseline classifier: Logistic Regression}

Logistic regression was used as a linear baseline model. The key settings were:

\begin{itemize}
    \item \texttt{penalty}: \texttt{l2}
    \item \texttt{solver}: \texttt{lbfgs}
    \item \texttt{max\_iter}: \texttt{1000}
    \item \texttt{C}: \texttt{1.0}
\end{itemize}

All other parameters remained at scikit-learn’s default values (version 1.7.0).

\subsection{Baseline classifier: Multi-layer Perceptron}

The MLP model consisted of a single hidden layer of 256 units and was trained using the Adam optimizer. The main configuration included:

\begin{itemize}
    \item \texttt{hidden\_layer\_sizes}: \texttt{256}
    \item \texttt{activation}: \texttt{relu}
    \item \texttt{batch\_size}: \texttt{512}
    \item \texttt{learning\_rate\_init}: \texttt{0.001}
    \item \texttt{solver}: \texttt{adam}
    \item \texttt{max\_iter}: \texttt{200}
\end{itemize}

Remaining hyperparameters used default values from the scikit-learn implementation.

\subsection{Main explainer: Shapley}

Model interpretability was assessed using SHAP (SHapley Additive exPlanations). We used the TreeExplainer from the \texttt{shap} library, which provides exact Shapley values for tree-based models such as XGBoost. The following settings were used:

\begin{itemize}
\item \textbf{Explainer:} \texttt{shap.TreeExplainer(model, data=x\_train)}
\item \textbf{Model output:} \texttt{"raw"} (default)
\item \textbf{Feature perturbation:} \texttt{"auto"} (uses \texttt{"tree\_path\_dependent"} for tree models)
\item \textbf{Approximate:} \texttt{False}
\item \textbf{Link function:} \texttt{None} (identity link)
\end{itemize}

SHAP values were computed on the training dataset, with each explainer requiring an average of 204 seconds per model-label pair.

\clearpage

\section{Feature comparison across binary outcomes}

\subsection{Respiratory}

Table \ref{tab:respiratory} summarizes key ECG feature comparisons between patients with and without respiratory cancer diagnoses across the MIMIC-IV and ECG-VIEW II cohorts. Across both cohorts, the positive samples against the negative shows consistent changes in ECG features such as an increase of P wave axis, decrease of QRS, decrease of QT, decrease of RR and increase of T wave axis.

\begin{table*}[ht!]
    \centering
    \caption{Comparison of ECG-derived features across patients with and without respiratory cancer diagnoses in MIMIC-IV and ECG-VIEW II cohorts. Diagnoses include overall lung cancer (ICD-10: C34), upper lobe lung cancer (C341), lower lobe lung cancer (C343), and unspecified part of lung cancer (C349). Feature distributions are summarized separately for positive and negative cases in each cohort.}
    \resizebox{\textwidth}{!}{%
    \begin{tabular}{llcccc}
    \hline
    \textbf{ICD-10 Code: Description} & \textbf{Feature} & \textbf{MIMIC-IV Negative} & \textbf{MIMIC-IV Positive} & \textbf{ECG-VIEW II Negative} & \textbf{ECG-VIEW II Positive} \\
    \hline\hline

    \multirow{8}{*}{C34: Lung cancer}
& RR & 769.00 (645.00, 909.00) & 659.00 (556.00, 779.00) & 857.00 (750.00, 968.00) & 759.00 (638.00, 882.00) \\
& PR & 158.00 (140.00, 178.00) & 150.00 (133.00, 169.00) & 158.00 (144.00, 172.00) & 154.00 (140.00, 170.00) \\
& QRS & 94.00 (86.00, 108.00) & 90.00 (82.00, 102.00) & 90.00 (84.00, 98.00) & 88.00 (82.00, 96.00) \\
& QT & 394.00 (360.00, 428.00) & 364.00 (330.00, 398.00) & 392.00 (368.00, 418.00) & 372.00 (344.00, 400.00) \\
& QTc & 447.66 (426.07, 473.43) & 445.84 (426.48, 468.81) & 421.00 (405.00, 442.00) & 424.00 (407.00, 445.00) \\
& P wave axis & 51.00 (32.00, 64.00) & 57.00 (41.00, 70.00) & 53.00 (37.00, 65.00) & 56.00 (41.00, 68.00) \\
& QRS axis & 13.00 (-15.00, 46.00) & 21.00 (-7.00, 54.00) & 48.00 (19.00, 68.00) & 47.00 (18.00, 67.00) \\
& T wave axis & 42.00 (13.00, 71.00) & 48.00 (18.00, 73.00) & 44.00 (27.00, 60.00) & 52.00 (34.00, 67.00) \\
    \hline

    \multirow{8}{*}{C341: Upper lung cancer}
& RR & 769.00 (645.00, 909.00) & 659.00 (566.00, 779.00) & 857.00 (741.00, 968.00) & 759.00 (632.00, 870.00) \\
& PR & 158.00 (140.00, 178.00) & 150.00 (132.00, 168.00) & 158.00 (144.00, 172.00) & 162.00 (146.00, 174.00) \\
& QRS & 94.00 (85.00, 108.00) & 90.00 (82.00, 102.00) & 90.00 (84.00, 98.00) & 86.00 (82.00, 96.00) \\
& QT & 394.00 (360.00, 428.00) & 366.00 (332.00, 398.00) & 392.00 (368.00, 416.00) & 366.00 (334.00, 392.00) \\
& QTc & 447.66 (426.09, 473.38) & 445.82 (426.08, 470.97) & 421.00 (405.00, 442.00) & 414.00 (402.00, 433.00) \\
& P wave axis & 51.00 (32.00, 64.00) & 57.00 (41.00, 70.00) & 53.00 (37.00, 65.00) & 54.00 (34.75, 66.00) \\
& QRS axis & 13.00 (-15.00, 46.00) & 21.50 (-5.00, 54.00) & 48.00 (19.00, 68.00) & 49.00 (26.00, 66.00) \\
& T wave axis & 42.00 (13.00, 71.00) & 49.00 (19.00, 74.00) & 44.00 (27.00, 60.00) & 49.00 (31.50, 62.00) \\
    \hline

    \multirow{8}{*}{C343: Lower lung cancer}
& RR & 769.00 (645.00, 909.00) & 666.00 (566.00, 800.00) & 857.00 (741.00, 968.00) & 750.00 (638.00, 879.00) \\
& PR & 158.00 (140.00, 178.00) & 150.00 (134.00, 174.00) & 158.00 (144.00, 172.00) & 152.00 (138.00, 168.50) \\
& QRS & 94.00 (85.00, 108.00) & 92.00 (84.00, 104.00) & 90.00 (84.00, 98.00) & 88.00 (82.00, 96.00) \\
& QT & 394.00 (360.00, 428.00) & 368.00 (334.00, 406.00) & 392.00 (368.00, 416.00) & 372.00 (350.00, 397.50) \\
& QTc & 447.66 (426.08, 473.39) & 445.82 (427.79, 468.31) & 421.00 (405.00, 442.00) & 427.50 (412.00, 447.00) \\
& P wave axis & 51.00 (32.00, 64.00) & 56.00 (36.00, 68.00) & 53.00 (37.00, 65.00) & 55.00 (41.00, 68.00) \\
& QRS axis & 13.00 (-15.00, 46.00) & 20.00 (-5.75, 54.00) & 48.00 (19.00, 68.00) & 41.00 (13.00, 66.50) \\
& T wave axis & 42.00 (13.00, 71.00) & 45.00 (11.00, 73.00) & 44.00 (27.00, 60.00) & 50.00 (31.50, 63.00) \\
    \hline

    \multirow{8}{*}{C349: Unspecified lung cancer}
& RR & 769.00 (645.00, 909.00) & 659.00 (560.00, 779.00) & 857.00 (750.00, 968.00) & 759.00 (638.00, 882.00) \\
& PR & 158.00 (140.00, 178.00) & 150.00 (134.00, 170.00) & 158.00 (144.00, 172.00) & 154.00 (140.00, 170.00) \\
& QRS & 94.00 (85.00, 108.00) & 90.00 (82.00, 102.00) & 90.00 (84.00, 98.00) & 88.00 (82.00, 96.00) \\
& QT & 394.00 (360.00, 428.00) & 364.00 (330.00, 399.00) & 392.00 (368.00, 418.00) & 372.00 (344.00, 400.00) \\
& QTc & 447.65 (426.08, 473.42) & 447.66 (427.24, 469.59) & 421.00 (405.00, 442.00) & 423.00 (406.00, 445.00) \\
& P wave axis & 51.00 (32.00, 64.00) & 58.00 (42.00, 71.00) & 53.00 (37.00, 65.00) & 56.00 (41.00, 68.00) \\
& QRS axis & 13.00 (-15.00, 46.00) & 19.00 (-12.00, 53.00) & 48.00 (19.00, 68.00) & 47.00 (18.00, 67.00) \\
& T wave axis & 42.00 (13.00, 71.00) & 48.00 (16.00, 73.00) & 44.00 (27.00, 60.00) & 52.00 (34.00, 67.00) \\
    \hline

    \end{tabular}
    }
    \label{tab:respiratory}
\end{table*}

\clearpage

\subsection{Urological}

Table \ref{tab:urological} summarizes key ECG feature comparisons between patients with and without urological cancer diagnoses across the MIMIC-IV and ECG-VIEW II cohorts. Across both cohorts, the positive samples against the negative shows consistent changes in ECG features such as an increase of PR, decrease of QRS axis, and increase of T wave axis.

\begin{table*}[ht!]
    \centering
    \caption{Comparison of ECG-derived features across patients with and without urological cancer diagnoses in MIMIC-IV and ECG-VIEW II cohorts. Diagnoses include prostate cancer (ICD-10: C61), BPH (N40), BPH w/o symptoms (N400), and bladder cancer (C679). Feature distributions are summarized separately for positive and negative cases in each cohort.}
    \resizebox{\textwidth}{!}{%
    \begin{tabular}{llcccc}
    \hline
    \textbf{ICD-10 Code: Description} & \textbf{Feature} & \textbf{MIMIC-IV Negative} & \textbf{MIMIC-IV Positive} & \textbf{ECG-VIEW II Negative} & \textbf{ECG-VIEW II Positive} \\
    \hline\hline

    \multirow{8}{*}{C61: Prostate cancer}
& RR & 769.00 (645.00, 909.00) & 759.00 (631.00, 896.00) & 857.00 (741.00, 968.00) & 870.00 (750.00, 984.00) \\
& PR & 158.00 (140.00, 178.00) & 162.00 (142.00, 186.00) & 158.00 (144.00, 172.00) & 166.00 (150.00, 182.00) \\
& QRS & 94.00 (85.00, 107.00) & 98.00 (88.00, 118.00) & 90.00 (84.00, 98.00) & 92.00 (86.00, 102.00) \\
& QT & 394.00 (360.00, 428.00) & 396.00 (358.00, 432.00) & 392.00 (368.00, 416.00) & 394.00 (370.00, 421.50) \\
& QTc & 447.55 (426.04, 473.34) & 454.66 (431.69, 481.27) & 421.00 (405.00, 442.00) & 420.00 (406.00, 442.00) \\
& P wave axis & 51.00 (32.00, 64.00) & 49.50 (27.00, 65.00) & 53.00 (37.00, 65.00) & 57.00 (41.00, 68.00) \\
& QRS axis & 14.00 (-15.00, 46.00) & 0.00 (-31.00, 33.00) & 48.00 (19.00, 68.00) & 39.00 (7.00, 61.00) \\
& T wave axis & 42.00 (13.00, 71.00) & 47.00 (13.00, 83.50) & 44.00 (27.00, 60.00) & 49.00 (31.00, 64.00) \\
    \hline

    \multirow{8}{*}{N40: BPH}
& RR & 769.00 (645.00, 909.00) & 779.00 (652.00, 923.00) & 857.00 (741.00, 968.00) & 845.00 (714.00, 968.00) \\
& PR & 158.00 (140.00, 178.00) & 166.00 (146.00, 192.00) & 158.00 (144.00, 172.00) & 166.00 (150.00, 184.00) \\
& QRS & 94.00 (84.00, 106.00) & 102.00 (90.00, 130.00) & 90.00 (84.00, 98.00) & 94.00 (86.00, 104.00) \\
& QT & 392.00 (360.00, 427.00) & 403.00 (366.00, 440.00) & 392.00 (368.00, 416.00) & 396.00 (370.00, 422.00) \\
& QTc & 447.03 (425.84, 472.84) & 455.63 (432.31, 483.74) & 421.00 (405.00, 442.00) & 432.00 (413.00, 453.00) \\
& P wave axis & 51.00 (33.00, 64.00) & 48.00 (25.00, 63.00) & 53.00 (37.00, 65.00) & 56.00 (40.00, 68.00) \\
& QRS axis & 15.00 (-13.00, 47.00) & 0.00 (-36.00, 33.00) & 48.00 (19.00, 68.00) & 32.00 (2.00, 60.00) \\
& T wave axis & 42.00 (14.00, 70.00) & 47.00 (10.00, 85.00) & 44.00 (27.00, 60.00) & 50.00 (31.00, 67.00) \\
    \hline

    \multirow{8}{*}{N400: BPH w/o symptoms}
& RR & 769.00 (645.00, 909.00) & 789.00 (659.00, 923.00) & 857.00 (741.00, 968.00) & 833.00 (709.50, 968.00) \\
& PR & 158.00 (140.00, 178.00) & 168.00 (146.00, 192.00) & 158.00 (144.00, 172.00) & 166.00 (150.00, 184.00) \\
& QRS & 94.00 (84.00, 106.00) & 102.00 (90.00, 130.00) & 90.00 (84.00, 98.00) & 94.00 (86.00, 104.00) \\
& QT & 392.00 (360.00, 427.00) & 405.00 (368.00, 440.00) & 392.00 (368.00, 416.00) & 396.00 (370.00, 422.00) \\
& QTc & 447.21 (426.00, 473.03) & 455.37 (432.00, 483.54) & 421.00 (405.00, 442.00) & 432.00 (413.00, 453.00) \\
& P wave axis & 51.00 (33.00, 64.00) & 47.00 (25.00, 63.00) & 53.00 (37.00, 65.00) & 56.00 (40.00, 68.00) \\
& QRS axis & 14.00 (-14.00, 47.00) & -1.00 (-36.00, 32.00) & 48.00 (19.00, 68.00) & 33.00 (1.00, 60.00) \\
& T wave axis & 42.00 (14.00, 70.00) & 47.00 (10.00, 85.00) & 44.00 (27.00, 60.00) & 50.00 (31.00, 67.00) \\
    \hline

    \multirow{8}{*}{C679: Bladder cancer}
& RR & 769.00 (645.00, 909.00) & 754.50 (618.00, 895.00) & 857.00 (741.00, 968.00) & 833.00 (723.00, 952.00) \\
& PR & 158.00 (140.00, 178.00) & 162.00 (140.00, 188.00) & 158.00 (144.00, 172.00) & 160.00 (146.00, 176.00) \\
& QRS & 94.00 (85.00, 108.00) & 96.00 (86.00, 126.50) & 90.00 (84.00, 98.00) & 90.00 (84.00, 100.00) \\
& QT & 394.00 (360.00, 428.00) & 390.00 (352.00, 430.00) & 392.00 (368.00, 416.00) & 388.00 (364.00, 412.00) \\
& QTc & 447.60 (426.08, 473.38) & 452.00 (428.21, 480.37) & 421.00 (405.00, 442.00) & 422.00 (405.00, 443.00) \\
& P wave axis & 51.00 (32.00, 64.00) & 50.50 (30.75, 64.00) & 53.00 (37.00, 65.00) & 57.00 (42.00, 69.00) \\
& QRS axis & 13.00 (-15.00, 46.00) & 3.00 (-25.00, 34.00) & 48.00 (19.00, 68.00) & 42.00 (13.00, 63.00) \\
& T wave axis & 42.00 (13.00, 71.00) & 54.00 (19.50, 87.00) & 44.00 (27.00, 60.00) & 50.00 (35.00, 65.00) \\
    \hline

    \end{tabular}
    }
    \label{tab:urological}
\end{table*}

\clearpage

\subsection{Digestive}

Table \ref{tab:gastro} summarizes key ECG feature comparisons between patients with and without digestive cancer diagnoses across the MIMIC-IV and ECG-VIEW II cohorts. Across both cohorts, the positive samples against the negative shows consistent changes in ECG features such as a decrease of PR, an increase of QTc, and a decrease of RR.

\begin{table*}[ht!]
    \centering
    \caption{Comparison of ECG-derived features across patients with and without digestive cancer diagnoses in MIMIC-IV and ECG-VIEW II cohorts. Diagnoses include esophageal cancer (ICD-10: C15), liver cancer (C22), and biliary tract cancer (C24). Feature distributions are summarized separately for positive and negative cases in each cohort.}
    \resizebox{\textwidth}{!}{%
    \begin{tabular}{llcccc}
    \hline
    \textbf{ICD-10 Code: Description} & \textbf{Feature} & \textbf{MIMIC-IV Negative} & \textbf{MIMIC-IV Positive} & \textbf{ECG-VIEW II Negative} & \textbf{ECG-VIEW II Positive} \\
    \hline\hline

    \multirow{8}{*}{C15: Esophageal cancer} 
& RR & 769.00 (645.00, 909.00) & 638.00 (545.00, 769.00) & 857.00 (741.00, 968.00) & 811.00 (674.00, 952.00) \\
& PR & 158.00 (140.00, 178.00) & 146.00 (128.00, 162.00) & 158.00 (144.00, 172.00) & 154.00 (140.00, 168.00) \\
& QRS & 94.00 (85.00, 108.00) & 90.00 (82.00, 102.00) & 90.00 (84.00, 98.00) & 88.00 (82.00, 96.00) \\
& QT & 394.00 (360.00, 428.00) & 362.50 (330.00, 398.25) & 392.00 (368.00, 416.00) & 386.00 (358.00, 414.00) \\
& QTc & 447.60 (426.08, 473.38) & 451.63 (427.96, 477.49) & 421.00 (405.00, 442.00) & 424.00 (408.00, 446.00) \\
& P wave axis & 51.00 (32.00, 64.00) & 53.00 (31.00, 66.00) & 53.00 (37.00, 65.00) & 63.00 (46.00, 74.00) \\
& QRS axis & 13.00 (-15.00, 46.00) & 12.00 (-14.00, 42.00) & 48.00 (19.00, 68.00) & 55.00 (28.00, 71.00) \\
& T wave axis & 42.00 (13.00, 71.00) & 34.00 (1.00, 62.00) & 44.00 (27.00, 60.00) & 57.00 (39.00, 70.00) \\
    \hline

    \multirow{8}{*}{C22: Liver cancer} 
& RR & 769.00 (645.00, 909.00) & 722.00 (600.00, 845.00) & 857.00 (741.00, 968.00) & 833.00 (706.00, 952.00) \\
& PR & 158.00 (140.00, 178.00) & 154.00 (138.00, 172.00) & 158.00 (144.00, 172.00) & 156.00 (142.00, 170.00) \\
& QRS & 94.00 (85.00, 108.00) & 92.00 (84.00, 102.00) & 90.00 (84.00, 98.00) & 90.00 (84.00, 98.00) \\
& QT & 394.00 (360.00, 428.00) & 384.00 (348.00, 421.00) & 392.00 (368.00, 416.00) & 394.00 (368.00, 420.00) \\
& QTc & 447.58 (426.04, 473.38) & 450.57 (431.49, 472.98) & 421.00 (405.00, 442.00) & 430.00 (410.00, 454.00) \\
& P wave axis & 51.00 (32.00, 64.00) & 47.00 (24.00, 62.00) & 53.00 (37.00, 65.00) & 52.00 (35.00, 66.00) \\
& QRS axis & 14.00 (-15.00, 46.00) & 9.00 (-13.00, 36.00) & 48.00 (19.00, 69.00) & 39.00 (13.00, 60.00) \\
& T wave axis & 42.00 (14.00, 71.00) & 32.00 (8.00, 55.75) & 44.00 (27.00, 61.00) & 42.00 (26.00, 58.00) \\
    \hline

    \multirow{8}{*}{C24: Biliary tract cancer} 
& RR & 769.00 (645.00, 909.00) & 714.00 (606.00, 845.00) & 857.00 (741.00, 968.00) & 800.00 (682.00, 923.00) \\
& PR & 158.00 (140.00, 178.00) & 150.00 (136.00, 166.00) & 158.00 (144.00, 172.00) & 154.00 (140.00, 168.00) \\
& QRS & 94.00 (85.00, 108.00) & 90.00 (84.00, 100.00) & 90.00 (84.00, 98.00) & 88.00 (82.00, 96.00) \\
& QT & 394.00 (360.00, 428.00) & 382.00 (344.00, 418.00) & 392.00 (368.00, 416.00) & 386.00 (360.00, 410.00) \\
& QTc & 447.65 (426.08, 473.38) & 448.46 (431.00, 470.56) & 421.00 (405.00, 442.00) & 425.00 (408.00, 447.00) \\
& P wave axis & 51.00 (32.00, 64.00) & 50.00 (29.00, 61.50) & 53.00 (37.00, 65.00) & 54.00 (38.00, 67.00) \\
& QRS axis & 13.00 (-15.00, 46.00) & -4.00 (-26.00, 25.00) & 48.00 (19.00, 68.00) & 40.00 (13.00, 60.00) \\
& T wave axis & 42.00 (13.00, 71.00) & 33.00 (5.00, 62.00) & 44.00 (27.00, 60.00) & 46.00 (27.00, 63.00) \\
    \hline

    \end{tabular}
    }
    \label{tab:gastro}
\end{table*}

\clearpage

\subsection{Gynecological}

Table \ref{tab:gyne} summarizes key ECG feature comparisons between patients with and without gynecological cancer diagnoses across the MIMIC-IV and ECG-VIEW II cohorts. Across both cohorts, the positive samples against the negative shows consistent changes in ECG features such as a decrease of PR, decrease of QRS, increase of QRS axis, decrease of QT, decrease of QTc and decrease of T wave axis.

\begin{table*}[ht!]
    \centering
    \caption{Comparison of ECG-derived features across patients with and without gynecological cancer diagnoses in MIMIC-IV and ECG-VIEW II cohorts. Diagnoses include uterine leiomyoma (ICD-10: D25), and endometriosis (N80). Feature distributions are summarized separately for positive and negative cases in each cohort.}
    \resizebox{\textwidth}{!}{%
    \begin{tabular}{llcccc}
    \hline
    \textbf{ICD-10 Code: Description} & \textbf{Feature} & \textbf{MIMIC-IV Negative} & \textbf{MIMIC-IV Positive} & \textbf{ECG-VIEW II Negative} & \textbf{ECG-VIEW II Positive} \\
    \hline\hline
    \multirow{8}{*}{D25: Uterine leiomyoma} 
& RR & 769.00 (645.00, 909.00) & 706.00 (594.00, 845.00) & 857.00 (741.00, 968.00) & 857.00 (769.00, 938.00) \\
& PR & 158.00 (140.00, 178.00) & 149.00 (134.00, 166.00) & 158.00 (144.00, 174.00) & 150.00 (138.00, 164.00) \\
& QRS & 94.00 (85.00, 108.00) & 86.00 (80.00, 94.00) & 90.00 (84.00, 98.00) & 86.00 (80.00, 92.00) \\
& QT & 394.00 (360.00, 428.00) & 376.00 (338.00, 410.00) & 392.00 (368.00, 416.00) & 390.00 (368.00, 410.00) \\
& QTc & 447.66 (426.09, 473.43) & 440.66 (423.99, 459.90) & 421.00 (405.00, 443.00) & 419.00 (406.00, 436.00) \\
& P wave axis & 51.00 (32.00, 64.00) & 52.00 (39.00, 63.00) & 53.00 (37.00, 65.00) & 52.00 (35.00, 64.00) \\
& QRS axis & 13.00 (-15.00, 46.00) & 23.00 (4.00, 45.00) & 47.00 (19.00, 68.00) & 55.00 (33.00, 70.00) \\
& T wave axis & 42.00 (13.00, 71.00) & 31.50 (10.00, 55.00) & 45.00 (27.00, 61.00) & 39.00 (26.00, 51.00) \\
    \hline
    \multirow{8}{*}{N80: Endometriosis} 
& RR & 769.00 (645.00, 909.00) & 718.00 (571.00, 833.00) & 857.00 (741.00, 968.00) & 845.00 (769.00, 938.00) \\
& PR & 158.00 (140.00, 178.00) & 146.00 (130.00, 162.00) & 158.00 (144.00, 174.00) & 148.00 (136.00, 160.00) \\
& QRS & 94.00 (85.00, 108.00) & 86.00 (80.00, 92.00) & 90.00 (84.00, 98.00) & 86.00 (80.00, 92.00) \\
& QT & 394.00 (360.00, 428.00) & 372.00 (334.00, 402.75) & 392.00 (368.00, 416.00) & 388.00 (368.00, 408.00) \\
& QTc & 447.66 (426.09, 473.38) & 436.76 (421.61, 455.49) & 421.00 (405.00, 443.00) & 418.00 (404.00, 436.00) \\
& P wave axis & 51.00 (32.00, 64.00) & 51.00 (33.50, 63.50) & 53.00 (37.00, 65.00) & 53.00 (37.00, 66.00) \\
& QRS axis & 13.00 (-15.00, 46.00) & 32.00 (12.00, 55.00) & 47.00 (19.00, 68.00) & 62.00 (40.00, 75.00) \\
& T wave axis & 42.00 (13.00, 71.00) & 31.00 (11.00, 48.00) & 44.00 (27.00, 61.00) & 42.00 (27.00, 53.00) \\
    \hline
    \end{tabular}
    }
    \label{tab:gyne}
\end{table*}

\subsection{Cerebral}

Table \ref{tab:cerebral} summarizes key ECG feature comparisons between patients with and without the cerebral cancer diagnose across the MIMIC-IV and ECG-VIEW II cohorts. Across both cohorts, the positive samples against the negative shows consistent changes in ECG features such as an increase of P wave axis, decrease of PR, decrease of QRS, increase of QRS axis, decrease of QT, increase of RR, and decrease of T wave axis.

\begin{table*}[ht!]
    \centering
    \caption{Comparison of ECG-derived features across patients with and without cerebral cancer diagnoses in MIMIC-IV and ECG-VIEW II cohorts. Diagnoses include brain metastases (ICD-10: C793). Feature distributions are summarized separately for positive and negative cases in each cohort.}
    \resizebox{\textwidth}{!}{%
    \begin{tabular}{llcccc}
    \hline
    \textbf{ICD-10 Code: Description} & \textbf{Feature} & \textbf{MIMIC-IV Negative} & \textbf{MIMIC-IV Positive} & \textbf{ECG-VIEW II Negative} & \textbf{ECG-VIEW II Positive} \\
    \hline\hline
    \multirow{8}{*}{C793: Brain metastases} 
& RR & 769.00 (645.00, 909.00) & 674.00 (566.00, 810.00) & 857.00 (741.00, 968.00) & 759.00 (652.00, 882.00) \\
& PR & 158.00 (140.00, 178.00) & 148.00 (132.00, 166.00) & 158.00 (144.00, 172.00) & 152.00 (136.00, 166.00) \\
& QRS & 94.00 (85.00, 108.00) & 88.00 (80.00, 98.00) & 90.00 (84.00, 98.00) & 86.00 (80.00, 94.00) \\
& QT & 394.00 (360.00, 428.00) & 364.00 (332.00, 400.00) & 392.00 (368.00, 416.00) & 376.00 (350.00, 400.00) \\
& QTc & 447.67 (426.13, 473.48) & 440.34 (421.72, 460.79) & 421.00 (405.00, 442.00) & 430.00 (411.00, 450.00) \\
& P wave axis & 51.00 (32.00, 64.00) & 54.00 (38.00, 66.00) & 53.00 (37.00, 65.00) & 56.00 (42.00, 68.00) \\
& QRS axis & 13.00 (-15.00, 46.00) & 21.00 (-5.00, 51.00) & 48.00 (19.00, 68.00) & 51.00 (21.00, 69.00) \\
& T wave axis & 42.00 (13.00, 71.00) & 45.00 (19.00, 67.00) & 44.00 (27.00, 60.00) & 47.00 (27.00, 64.00) \\
    \hline
    \end{tabular}
    }
    \label{tab:cerebral}
\end{table*}

\clearpage

\section{Benchmarking models}
Tab.~\ref{tab:benchmarking} compares three different model architectures in terms of predictive performance on the internal and external test set. We assess the model performance based on the following scheme: A model that performs best or stays consistent with the best-performing model for a task on both the internal and the external test set is flagged as \colgreen{green}. A model that performs best or remains consistent with the best-performing model on either the internal or the external test set is marked in \colyellow{orange}, otherwise in \colred{red}. The gradient-boosted decision tree (XGBoost) reaches 6 \colgreen{green}, 6 \colyellow{orange} and 2 \colred{red} scores. Logistic regression scores 4 \colgreen{green}, 6 \colyellow{orange} and 4 \colred{red}. Finally, the multi-layer perceptron reaches 3 \colgreen{green}, 4 \colyellow{orange} and 5 \colred{red}. These results underline that the three models show in many cases comparable performance. In order to reduce the complexity of the study, we focus in the main text on the results obtained for XGBoost, which shows the strongest overall performance across all prediction tasks. XGBoost and LR also compare favorably in comparison to MLP in terms of runtime and show further advantages in terms of explainablity.

\begin{table*}[ht!]
    \centering
    \caption{Comparison of internal and external AUROC with 95\% confidence intervals (CI) and computational time (in seconds) across classifiers for selected neoplasms. Time includes both training and evaluation. Best-performing model per condition is highlighted in bold face and underlined. We also mark results in bold face where the point prediction falls into the confidence interval of the best-performing model.}
    \label{tab:benchmarking}
    
    \resizebox{\textwidth}{!}{%
    \begin{tabular}{llccc}
    \hline
    \textbf{ICD-10 Code: Description} & \textbf{Model} & \textbf{Internal AUROC (95\% CI)} & \textbf{External AUROC (95\% CI)} & \textbf{Time (s)} \\ \hline\hline

    C34: Lung cancer & \colgreen{XGBoost} & \underline{\textbf{0.800}} (0.784, 0.815) & \underline{\textbf{0.767}} (0.765, 0.771) & \underline{\textbf{1215.8}} \\
                     & \colred{LR}      & 0.747 (0.720, 0.757) & 0.759 (0.755, 0.762) & 1237.2 \\
                     & \colred{MLP}     & 0.778 (0.775, 0.797) & 0.755 (0.754, 0.758) & 3834.7 \\ \hline

    C341: Upper lung cancer & \colgreen{XGBoost} & \underline{\textbf{0.723}} (0.706, 0.750) & \textbf{0.738} (0.723, 0.751) & \underline{\textbf{1209.8}} \\
                            & \colyellow{LR}      & 0.696 (0.645, 0.731) & \underline{\textbf{0.746}} (0.733, 0.767) & 1239.1 \\
                            & \colred{MLP}     & 0.700 (0.670, 0.751) & 0.720 (0.692, 0.735) & 3109.2 \\ \hline

    C343: Lower lung cancer & \colyellow{XGBoost} & \textbf{0.855} (0.788, 0.887) & 0.752 (0.747, 0.770) & \underline{\textbf{1154.4}} \\
                            & \colyellow{LR}      & 0.803 (0.777, 0.831) & \underline{\textbf{0.779}} (0.759, 0.795) & 1228.2 \\
                            & \colyellow{MLP}     & \underline{\textbf{0.863}} (0.817, 0.906) & 0.749 (0.734, 0.768) & 3307.9 \\ \hline

    C349: Unspecified lung cancer & \colgreen{XGBoost} & \underline{\textbf{0.792}} (0.755, 0.822) & \underline{\textbf{0.759}} (0.757, 0.760) & \underline{\textbf{1178.1}} \\
                                  & \colred{LR}      & 0.726 (0.704, 0.753) & 0.750 (0.746, 0.751) & 1226.3 \\
                                  & \colyellow{MLP}     & \textbf{0.777} (0.761, 0.810) & 0.755 (0.751, 0.757) & 3182.2 \\ \hline\hline

    C61: Prostate cancer & \colred{XGBoost} & 0.756 (0.746, 0.781) & 0.795 (0.792, 0.797) & 1298.1   \\
                         & \colgreen{LR}      & \underline{\textbf{0.803}} (0.778, 0.817) & \underline{\textbf{0.806}} (0.800, 0.809) & \underline{\textbf{600.8}}  \\
                         & \colgreen{MLP}     & \textbf{0.782} (0.766, 0.796) & \textbf{0.804} (0.800, 0.808) & 1865.5 \\ \hline

    N40: Benign prostatic hyperplasia (BPH) & \colgreen{XGBoost} & \underline{\textbf{0.749}} (0.739, 0.760) & \underline{\textbf{0.820}} (0.817, 0.821) & 1301.6   \\
                                            & \colgreen{LR}      & \underline{\textbf{0.749}} (0.744, 0.760) & \underline{\textbf{0.820}} (0.817, 0.824) & \underline{\textbf{583.4}}  \\
                                            & \colred{MLP}     & 0.736 (0.726, 0.743) & 0.815 (0.812, 0.817) & 9541.0 \\ \hline

    N400: BPH without symptoms & \colred{XGBoost} & 0.739 (0.727, 0.751) & 0.828 (0.823, 0.831) & 1164.1   \\
                               & \colgreen{LR}      & \underline{\textbf{0.743}} (0.740, 0.751) & \underline{\textbf{0.832}} (0.829, 0.834) & \underline{\textbf{589.7}}  \\
                               & \colred{MLP}     & 0.725 (0.716, 0.739) & 0.822 (0.817, 0.826) & 6558.4 \\ \hline

    C679: Bladder cancer, unspecified & \colyellow{XGBoost} & \textbf{0.833} (0.803, 0.893) & 0.757 (0.748, 0.762) & 1272.1   \\
                                      & \colgreen{LR}      & \underline{\textbf{0.852}} (0.817, 0.875) & \underline{\textbf{0.796}} (0.792, 0.801) & \underline{\textbf{1214.3}} \\
                                      & \colred{MLP}     & 0.784 (0.742, 0.820) & 0.781 (0.773, 0.788) & 3372.5 \\ \hline\hline

    C15: Esophageal cancer & \colyellow{XGBoost} & \textbf{0.818} (0.780, 0.876) & 0.810 (0.807, 0.815) & 1218.1 \\
                           & \colyellow{LR}      & \underline{\textbf{0.848}} (0.816, 0.874) & 0.774 (0.765, 0.779) & \underline{\textbf{1189.7}} \\
                           & \colyellow{MLP}     & 0.803 (0.794, 0.857) & \underline{\textbf{0.819}} (0.817, 0.825) & 3169.9 \\ \hline

    C22: Liver cancer & \colyellow{XGBoost} & \textbf{0.808} (0.782, 0.825) & 0.719 (0.715, 0.722) & \underline{\textbf{1151.7}} \\
                      & \colred{LR}      & 0.731 (0.687, 0.763) & 0.679 (0.677, 0.684) & 1225.5 \\
                      & \colgreen{MLP}     & \underline{\textbf{0.827}} (0.793, 0.844) & \underline{\textbf{0.739}} (0.734, 0.740) & 3033.6 \\ \hline

    C24: Biliary tract cancer & \colyellow{XGBoost} & \underline{\textbf{0.837}} (0.756, 0.904) & 0.706 (0.702, 0.712) & \underline{\textbf{1201.1}} \\
                              & \colyellow{LR}      & 0.713 (0.637, 0.784) & \textbf{0.730} (0.723, 0.741) & 1234.9 \\
                              & \colyellow{MLP}     & 0.740 (0.623, 0.857) & \underline{\textbf{0.734}} (0.726, 0.737) & 3076.5 \\ \hline\hline

    D25: Leiomyoma of uterus & \colgreen{XGBoost} & \underline{\textbf{0.808}} (0.735, 0.854) & \underline{\textbf{0.730}} (0.727, 0.732) & 1232.5   \\
                             & \colyellow{LR}      & \textbf{0.738} (0.669, 0.806) & 0.668 (0.665, 0.669) & \underline{\textbf{560.2}}  \\
                             & \colyellow{MLP}     & \textbf{0.798} (0.750, 0.824) & 0.713 (0.712, 0.715) & 1665.0 \\ \hline

    N80: Endometriosis        & \colgreen{XGBoost} & \underline{\textbf{0.879}} (0.845, 0.907) & \underline{\textbf{0.753}} (0.750, 0.757) & 1199.7   \\
                             & \colyellow{LR}      & \textbf{0.857} (0.841, 0.873) & 0.742 (0.739, 0.743) & \underline{\textbf{559.8}}  \\
                             & \colyellow{MLP}     & \textbf{0.876} (0.837, 0.905) & 0.745 (0.740, 0.748) & 1649.0 \\ \hline\hline

    C793: Brain metastases & \colyellow{XGBoost} & 0.738 (0.712, 0.762) & \textbf{0.699} (0.693, 0.707) & \underline{\textbf{1199.4}} \\
                           & \colred{LR}      & 0.740 (0.714, 0.753) & 0.654 (0.643, 0.665) & 1212.9 \\
                           & \colgreen{MLP}     & \underline{\textbf{0.768}} (0.754, 0.798) & \underline{\textbf{0.708}} (0.697, 0.719) & 3612.5 \\ \hline
    \end{tabular}
    }
    \label{tab:baselines_comparison}
\end{table*}

\end{document}